\documentclass[structabstract]{aa}
\usepackage{natbib}
\usepackage{graphicx}
\usepackage{txfonts}
\def\arcsec{\hbox{$^{\prime\prime}$}}

\def\lya{Ly$-\alpha$}
\def\lyb{Ly$-\beta$}

\begin{document}
   \title{Hydrogen Lyman$-\alpha$ and Lyman$-\beta$ spectral radiance profiles in the quiet Sun}
   \author{H. Tian\inst{1, 2}
          \and W. Curdt\inst{1}
          \and E. Marsch\inst{1}
          \and U. Sch\"uhle\inst{1}}
   \institute{Max-Planck-Institut f\"ur Sonnensystemforschung,
   Max-Planck-Str. 2, 37191 Katlenburg-Lindau, Germany\\
   \email{tianhui924@gmail.com}
   \and School of Earth and Space Sciences, Peking University, China}
\date{}

\abstract
   {}
   {We extend earlier work by studying in the quiet Sun the line profiles of
   the hydrogen Lyman-$\alpha$ and Lyman-$\beta$ lines, which were obtained
   quasi-simultaneously in a raster scan with a size of about
   150\arcsec~$\times$~120{\arcsec} near disk center.}
   {The self-reversal depths of the \lya~and \lyb~profiles are quantified
   by measuring the maximum spectral radiances of the two horns and the minimum
   spectral radiance of the central reversal. The information on the asymmetries
   of the \lya~and \lyb~profiles is obtained through the calculation of the 1$^{st}$ and
   3$^{rd}$-order moments of the line profiles. By comparing maps of self-reversal depths and
   the moments with radiance images of the Lyman lines, photospheric
   magnetograms, and Dopplergrams of two other optically thin lines, we study
   the spatial distribution of the \lya~and \lyb~profiles with different
   self-reversal depths, and investigate the relationship between profile
   asymmetries and flows in the solar atmosphere.}
   {We find that the emissions of the Lyman lines tend to be more strongly absorbed
   in the internetwork, as compared to those in the network region. Almost all of the
   \lya~ profiles are self-reversed, whilst about 17\% of the \lyb~profiles are not reversed. The
   ratio of \lya~and \lyb~ intensities seems to be independent of
   the magnetic field strength. Most \lya~profiles are stronger in the blue horn, while
   most \lyb~profiles are stronger in the red horn. However, the
   opposite asymmetries of \lya~and \lyb~are not pixel-to-pixel correlated.
   We also confirm that when larger transition-region downflows are present, the
   \lya~and \lyb~ profiles are more enhanced in the blue and red horns, respectively. The
   first-order moment of \lyb, which reflects the combined effects of the profile
   asymmetry and motion of the emitting material, strongly correlates with
   the Doppler shifts of the Si\,{\sc{iii}} and O\,{\sc{vi}} lines, whilst for
   \lya~this correlation is much weaker. Our analysis shows that both \lya~and \lyb~might be more
   redshifted if larger transition-region downflows are present. We also find that the observed average
   \lyb~profile is redshifted with respect to its rest position.}
   {}

    \titlerunning{Hydrogen \lya~and \lyb~profiles in the quiet Sun}
    \authorrunning{H. Tian et al.}
   \maketitle
%
\section{Introduction}
Hydrogen is the most abundant element in the solar atmosphere. And
thus its resonance lines, especially the Lyman-alpha (\lya) and
Lyman-beta (\lyb) lines, play an important role in the overall
radiative energy transport of the Sun \citep{Fontenla88}.

\lya~is by far the strongest line in the vacuum ultraviolet (VUV)
spectral range. It is so dominant that $\approx$75\% of the
integrated radiance from 800~{\AA} to 1500~{\AA} in the quiet Sun
come from this single line \citep{Wilhelm98}. Energy losses through
the \lya~ emission are the most important radiative losses in the
lower transition region, where the approximate temperature ranges
from 8000~K to 30000~K \citep{Fontenla88}. Also, the spectral
irradiance at the center of the solar \lya~line profile is the main
excitation source responsible for the atomic hydrogen resonant
scattering in cool cometary and planetary material, and thus is
required in modeling the \lya~emissions occurring in cool
interplanetary environments \citep{Emerich05}.

About three decades ago, detailed observations of the \lya~and \lyb~
line profiles were performed with the NRL slit spectrometer onboard
{\it Skylab} \citep[e.g.][]{Nicolas76} and the UV polychromator
onboard {\it OSO 8} (Orbiting Solar Observatory)
\citep[e.g.][]{Lemaire78,Vial82}. \citet{Basri79} used
high-resolution \lya~spectra obtained by the HRTS (High Resolution
Telescope and Spectrograph) instrument on rocket flight to
investigate the network contrast and center-to-limb variation in the
line profiles. Later, \lya~profiles with both high spectral and
spatial resolutions were obtained by the UVSP instrument onboard
{\it SMM} \citep[Solar Maximum Mission, e.g.,][]{Fontenla88}. The
early studies on the \lya~and \lyb~line profiles revealed clearly
that the Lyman lines are highly variable not only in time, but also
in space. However, these early observations were hampered by the
strong geocoronal absorption at the line center.

The problem of geocoronal absorption went away, and high spectral,
temporal, and spatial resolution observation of the Lyman lines
could be obtained, when in 1995 the {\it SoHO} (Solar and
Heliospheric observatory) space probe was positioned into an orbit
around the first Lagrangian Point, L$_1$. The Solar Ultraviolet
Measurements of Emitted Radiation spectrograph
\citep[SUMER,][]{Wilhelm95,Lemaire97} covers the whole hydrogen
Lyman series as well as the Lyman continuum \citep{Curdt01}, and
provides full line profiles in high spatial and spectral resolution
of ~1{\arcsec} and 44~m{\AA}, respectively. However, since the \lya~
line is so prominent, its high radiance leads to saturation of the
detector microchannel plates. To overcome this shortcoming, 1:10
attenuating grids above 50 pixels on both sides of the detectors had
been introduced into the optical design of the instrument
\citep{Wilhelm95}. But unfortunately, the attenuators also cause
unpredictable modifications of the line profiles. By using SUMER
data, \citet{Warren98} completed a comprehensive analysis on the
profiles of the higher H Lyman series lines (from \lyb~to
{Ly$-\lambda$\,($n=2,\cdots, 11$)}). They found that: (1) the
average profiles for \lyb~through {Ly$-\epsilon$\,($n$=5)} are
self-reversed, and the remaining lines are flat-topped; (2) the
network profiles show a strong enhancement in the red wings, while
the internetwork profiles are nearly symmetric; (3) the limb
brightening is weak. Higher Lyman lines obtained near the limb were
analysed by \citet{Marsch99} and \citet{Marsch00}. These authors
found that the line width of the Lyman lines increases with
decreasing main quantum number. \citet{Schmieder99} and
\citet{Heinzel01} presented a nearly simultaneous observation of all
the hydrogen Lyman lines including the \lya~line recorded on the
attenuator of SUMER. However, the line profile was distorted by the
already mentioned problem of the attenuator.

Several attempts were then made to observe \lya~on the bare part of
the SUMER detector \citep{Teriaca5a,Teriaca5b,Teriaca06}. These
authors extrapolated the gain-depression correction to the high
photon input rate attained during those exposures, which introduced
a high uncertainty in the signal determination. Using the scattered
light from the primary mirror, \citet{Lemaire98} deduced a nearly
symmetric \lya~profile of the full-Sun irradiance. Later,
\citet{Lemaire02} compared the profiles of \lya~and \lyb~and
reported calibrated irradiances with 10\% uncertainty. The
relationship between the \lya~line center irradiance and the total
line irradiance was also studied by using this kind of observation
\citep{Lemaire05,Emerich05}.

\citet{Xia03} studied the difference of \lyb~profiles in the coronal
hole and in the quiet Sun. He found that the asymmetry of the
average \lyb~line profile - the red-horn dominance - is stronger in
the quiet Sun than in the coronal hole. He also found more locations
with blue-horn dominance in \lyb~profiles in coronal holes than in
quiet-Sun regions. By fitting the two wings of the \lyb~profiles,
\citet{Xia04} derived the Doppler shift of the line, which was found
to have a correlation with the Doppler shift of the typical
transition region lines C\,{\sc{ii}} and O\,{\sc{vi}}.

Hydrogen Lyman lines were frequently used to reveal information on
the fine structures and physical properties of quiescent solar
prominence. Since different Lyman lines and their line center, peak,
and wings are formed at different depths within the prominence
thread, the Lyman series are important to diagnose the variation of
the thermodynamic conditions from the prominence-corona transition
region (PCTR) to the central cool parts \citep{Vial07}. To explain
the properties of observed Lyman line profiles, multi-thread
prominence fine-structure models consisting of 2D threads with
randomly assigned line-of-sight (LOS) velocities were developed
\citep{Gunar08}. With the assumption of magnetohydrostatic (MHS)
equilibrium, the models are based on an empirical PCTR and use
calculations of non-LTE (local thermodynamic equilibrium) radiative
transfer. Such models have shown that the profiles of Lyman lines
higher than \lya~are more reversed when seen across than along the
magnetic field lines \citep{Heinzel05}. This behaviour was confirmed
in a prominence observation by \citet{Schmieder07}.

A recent study showed that a LOS velocity of the order of 10~km/s
can lead to substantial asymmetries of the synthetic line profiles
obtained by the multi-thread modelling \citep{Gunar08}. This work
also predicted that the \lya~profiles may exhibit an asymmetry
opposite to those of higher Lyman lines. The ratio of \lya/\lyb~is
also very sensitive to the physical and geometrical properties of
the fine structures in prominences, and the fluctuations of this
ratio are believed to relate to such fine structures \citep{Vial07}.
The {\it OSO}~8 observation yielded a value of 65 (in the energy
unit) for this ratio \citep{Vial82}, while a recent SUMER
observation revealed a different ratio (96, 183, and 181, in the
energy unit) in different parts of a prominence \citep{Vial07}.

The \lya~and \lyb~line profiles can also be used to diagnose
nonthermal effects in solar flares, e.g. the nonthermal ionization
of hydrogen by electron beams. It is predicted that, at the early
stage of the impulsive phase of flares, the nonthermal effect should
be strong and the \lya~and \lyb~profiles will be broadened and
enhanced, especially in their wings; and after the maximum of the
impulsive phase, the intensity will decrease rapidly due to a rapid
increase of the coronal column mass \citep{Henoux95,Fang95}.

The Lyman line profiles observed in sunspot regions appear to be
different from those in the quiet Sun. Observations have shown that
the self-reversals are almost absent in sunspot regions
\citep{Fontenla88,Curdt01,Tian09}. In contrast, the lower Lyman line
profiles observed in the plage region are obviously reversed, a
phenomenon also found in the normal quiet Sun \citep{Tian09}. These
results indicate a much smaller opacity above sunspots, as compared
to the surrounding plage region.

\begin{figure*}
   \sidecaption
   \includegraphics[width=12.5cm]{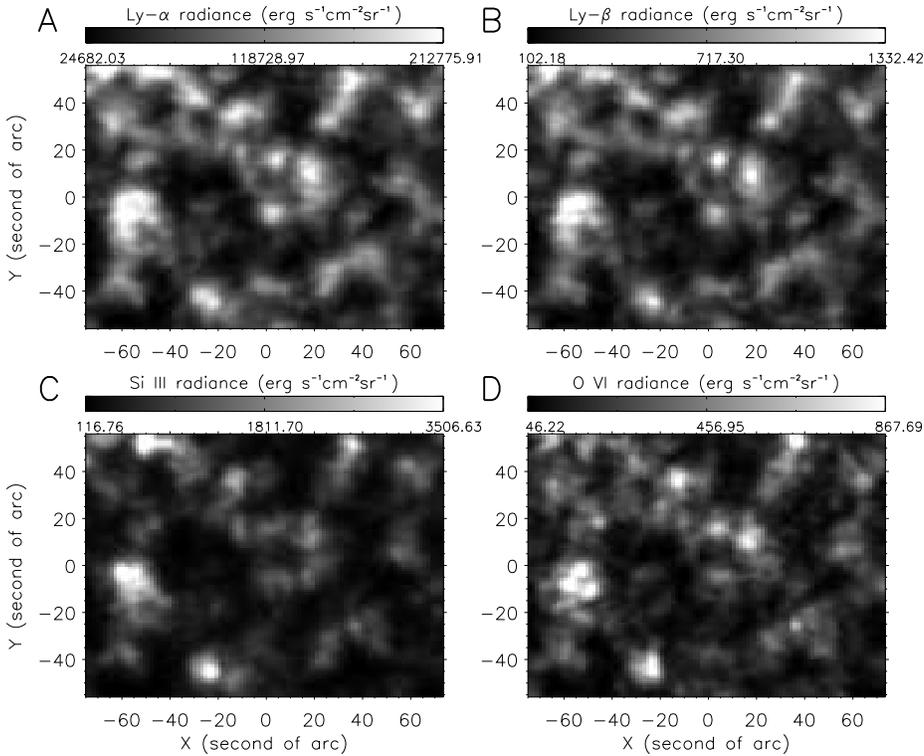}
   \caption{Radiance maps of two Lyman and two transition-region lines:
   (A) \lya~($\lambda$\,1216), (B) \lyb~($\lambda$\,1025), (C)  Si\,{\sc{iii}}
   ($\lambda$\,1206), and (D) O\,{\sc{vi}} ($\lambda$\,1032).}
   \label{fig.1}
\end{figure*}

It was proposed that the asymmetries of the Lyman lines are probably
related to large-scale motions of the atmosphere
\citep{Gouttebroze78}. In the energy-balance model of the
chromosphere-corona transition region \citep{Fontenla02}, the
authors concluded that the H and He line profiles are greatly
affected by flows, while in a recent multi-thread prominence model
it was suggested that asymmetrical line profiles are produced by the
combined effects of different Doppler shifts and absorption
coefficients (optical thicknesses) in the individual threads
\citep{Gunar08}. The authors also suggested that opposite
asymmetries in the profiles of \lya~and higher Lyman lines are
probably caused by different line opacities. However, much further
work is still needed to understand the origin of the line
asymmetries.

In a previous paper \citep{Curdt08}, we presented results of a
non-routine observation sequence, in which we partly closed the
aperture door of SUMER to reduce the incoming photon flux to a
20\%-level and obtained high spectral and spatial resolution
profiles of \lya~and \lyb. We have shown that the averaged profiles
of \lya~and \lyb~exhibit opposite asymmetries, and that the
asymmetries depend on the Doppler flows in the transition region.
However, in that observation, the profiles of \lya~and \lyb~were
obtained in a sit-and-stare mode, and thus only a small vertical
slice of the Sun was sampled.

In this paper, we will present results of a more recent observation,
in which we scanned a rectangular region around the quiet Sun disk
center and obtained profiles of \lya~and \lyb~quasi-simultaneously.
This is a unique data set, since such an observation was never
carried out before. We find that the spatial distribution of
profiles with different self-reversals is correlated with the
network pattern, and that the dependence of the profile asymmetries
on the transition-region Doppler flows is different for \lya~and
\lyb.

\section{Observation and data reduction}

Motivated by the results obtained recently from a sit-and-stare
observation \citep[for details cf.,][]{Curdt08}, we modified the
observing sequence and rastered a rectangular region near the solar
disk center in order to get a better selection of solar features in
the data set. The high quality of this observation allows the
detailed statistical analysis presented here.

The SUMER observation was made at the center part of the solar disk
from 21:59 to 23:18 on September 23, 2008. Similarly to the
observation on July 2, 2008 \citep{Curdt08}, we partly closed the
aperture door -- during a real-time commanding session prior to the
observation and re-opening thereafter -- and could thus reduce the
input photon rate by a factor of $\approx5$. As a prologue to the
observation, full-detector images in the Lyman continuum around
{880~\AA} were obtained with open and partially-closed door. At this
wavelength setting, the illumination of the detector is rather
uniform, and no saturation effects had to be considered. In this
way, accurate values for the photon flux reduction could be
established. We found that the incoming photon flux was reduced to a
18.3\% level.

After this prologue, the slit~7 (0.3\arcsec~$\times$~120{\arcsec})
was used to scan a rectangular region with a size of about
150\arcsec~$\times$~120{\arcsec} at the disk center. We had to use
this narrow slit to avoid saturation of the detector. We completed
the raster with an increment of 1.5{\arcsec}, a value which is
comparable to the instrument point spread function (instrumental
resolution), so that undersampling effects should be minor. For each
spectral setting two spectral windows were transmitted to the
ground. For the \lya~setting, 100~pixels (px) around the
$\lambda$\,1216~H\,{\sc i} line were recorded on the bare
photocathode of the detector, and 50~px around the
$\lambda$\,1206~Si\,{\sc iii} line on the KBr part of the
photocathode, respectively; for the \lyb~setting, 100~px around the
$\lambda$\,1025~H\,{\sc i} line and 50~px around the
$\lambda$\,1032~O\,{\sc vi} line, were recorded both on the
KBr-coated section. With an exposure time of 15~s, all lines were
recorded on detector~B, with sufficient counts for a good
line-profile analysis. Each time, the back-and-forth movement of the
wavelength mechanism between subsequent exposures lasted for
$\approx10$~s and led to a cadence of 50.5~s. The observed region on
the Sun was very quiet during our observations.

The standard procedures for correcting and calibrating the SUMER
data were applied, including local-gain correction, dead-time
correction, flat-field correction, destretching, and radiometric
calibration. To complete the flat-field correction, we used the
flat-field exposure acquired on June 28, 2008. It turned out that
there is a small shift of this flat field relative to the actual
data. Therefore we shifted the correction matrix by a fraction of a
pixel (0.3 pixel in solar-X, and 0.7 pixel in solar-Y) before
completing the correction.

In order to improve the statistics, we averaged the nine profiles in
a square of 3~$\times$~3~pixels centered at each spatial pixel. This
process is like a running average of the profiles in both spatial
dimensions and will definitely increase the signal-to-noise ratio.
Then the radiances of the observed spectra were divided by the
factor of the photon flux reduction, which was 18.3\% in this
observation. The line radiance of each profile was obtained through
an integration of the line profile. The calibrated radiance maps of
the four lines are presented in Fig.~\ref{fig.1}.

During the SUMER observation, high-rate magnetograms were also
obtained by the Michelson Doppler Imager \citep[MDI,][]{Scherrer95}
onboard SOHO. The magnetograms have a pixel size of about
0.6~\arcsec~and reveal clearly the magnetic fluxes in the network.
In order to find a possible correlation between the magnetic flux
and the \lya~and \lyb~profiles, we selected nine magnetograms
observed from 22:06 to 22:15 UT and averaged them to increase the
signal-to-noise ratio.

\section{The self-reversal of the profiles}

\begin{figure*}
\sidecaption
\includegraphics[width=12.5cm]{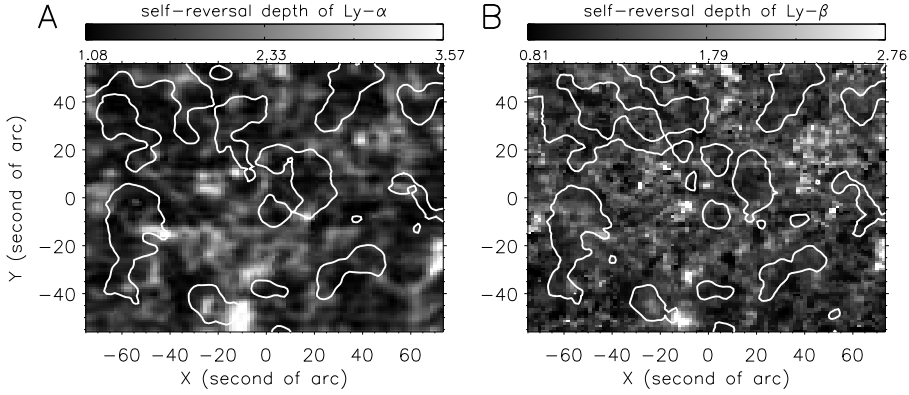}
\caption{Maps of the self-reversal depth (represented by $R_{br2c}$)
of the \lya~ and \lyb~line profiles, respectively. The white
contours (enclosing the top 25\% values) represent the brightest
parts on the corresponding radiance maps A and B of
Fig.~\ref{fig.1}.} \label{fig.2}
\end{figure*}

\begin{figure*}
\sidecaption
\resizebox{\hsize}{!}{\includegraphics{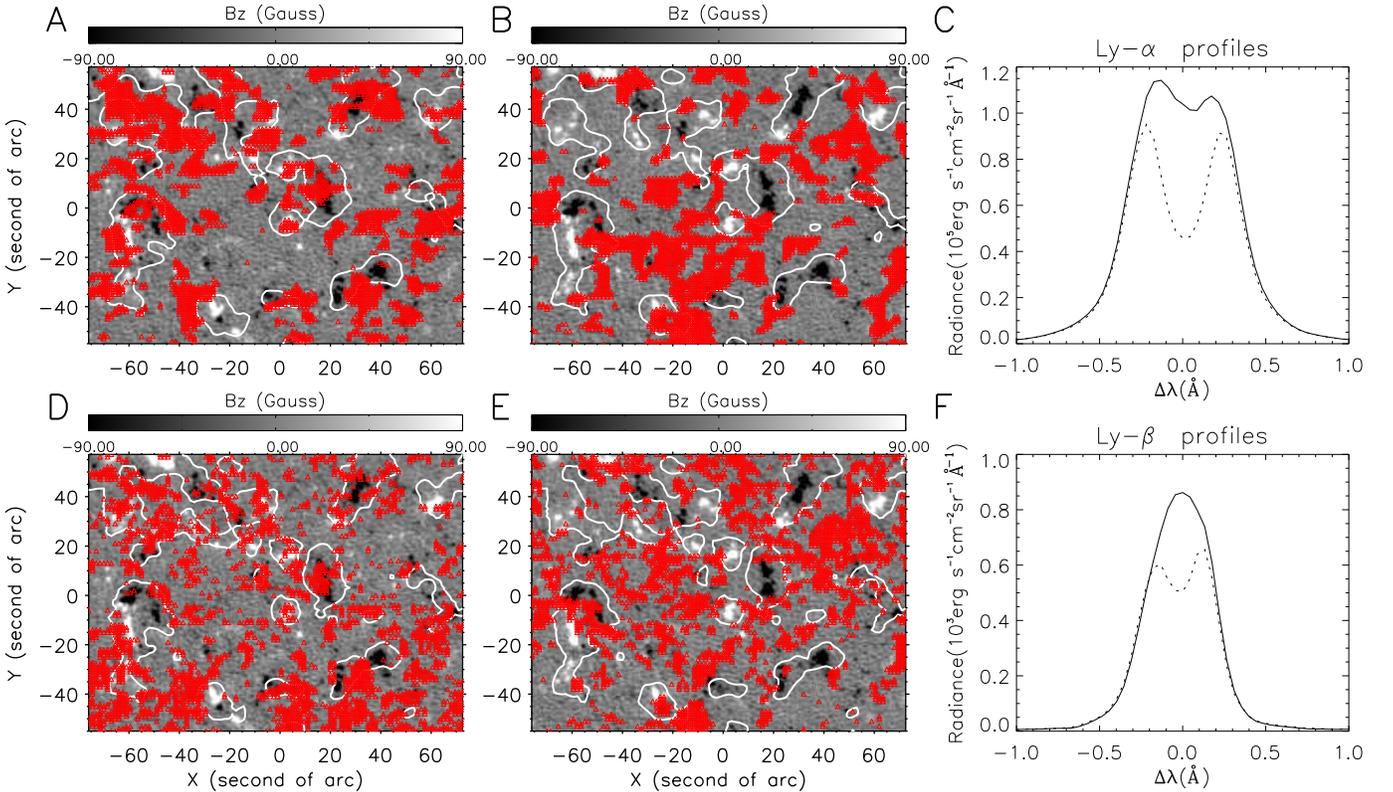}} \caption{Positions
of \lya~(upper panels A and B) and \lyb~(lower panels D and E)
profiles having shallow (left panels, corresponding to the 25\%
smallest values of $R_{br2c}$) or deep (central panels,
corresponding to the 25\% largest values of $R_{br2c}$)
self-reversals. The locations are marked by red triangles on the
grey-scale MDI magnetogram. The white contours (enclosing the top
25\% values) represent the brightest part on the corresponding
radiance map. The averaged \lya~ profiles at positions marked in
panel A/B are shown as the solid/dotted line in panel C. And the
averaged \lyb~ profiles at positions marked in panel D/E are shown
as the solid/dotted line in panel F. } \label{fig.3}
\end{figure*}

In the normal quiet Sun region, most \lya~and \lyb~ line profiles
exhibit a self-reversal at their centers \citep{Warren98,Curdt08}.
The depth of the self-reversal is a measure of the degree to which
the Lyman line emission is absorbed by the hydrogen atoms in upper
layers.

We first measured the maximum spectral radiances of the two horns
(the selected spectral range is 0.12~{\AA}-0.35~{\AA} from the line
center for \lya~and 0.09~{\AA}-0.24~{\AA} from the line center for
\lyb) and the minimum spectral radiance of the central part (within
0.12~{\AA} from the line center for \lya~and 0.09~{\AA} from the
line center for \lyb). The resulting spectral radiances of the
central part, blue and red horns are designated by $I_c$, $I_b$ and
$I_r$, respectively. By comparing the horn radiances with the
central radiance, we find that almost all of the \lya~profiles are
obviously self-reversed (only 3 profiles are flat-topped), whilst
about 17\% of the \lyb~profiles are flat-topped or not reversed. For
the non-reversed profiles, $I_c$ was calculated as the mean value of
the spectral radiance of the central part, instead of the minimum
value. The extent of the self-reversal is an indicator of the
opacity and can be quantified by the ratio of the average horn
intensity relative to the central intensity, which is designated by
$R_{br2c}=0.5(I_b+I_r)/I_c$. Although this method might not be very
accurate for some profiles which are too noisy or have no obvious
central reversal and horns, in a statistical sense the parameter
$R_{br2c}$ should be a good candidate to represent the relative
depth of the self-reversal of different profiles. After calculating
the value of $R_{br2c}$ for each profile, we constructed maps of the
self-reversal depths of \lya~and \lyb~ as shown in Fig.~\ref{fig.2}.
The white contours (enclosing the top 25\% values) on the maps
represent the brightest parts on the corresponding radiance maps of
Fig.~\ref{fig.1}, which clearly outline the network pattern.

The self-reversal depths of both \lya~and \lyb~profiles shown in
Fig.~\ref{fig.2} are statistically smaller in the network than in
the internetwork, which indicates that the internetwork emissions
are subject to stronger absorption than those in the network. This
behaviour is confirmed in Fig.~\ref{fig.3}, in which the positions
of weakly (corresponding to the 25\% smallest values of $R_{br2c}$)
or strongly (corresponding to the 25\% largest values of $R_{br2c}$)
absorbed emissions are marked with red triangles. The background
image is an MDI magnetogram, and the contours are the same as those
in Fig.~\ref{fig.2}. The averaged profiles with shallow and deep
reversals are presented in the right panels, which clearly reveals
the difference. From Figures~\ref{fig.3}A and D we find that the
locations of some weakly absorbed emissions deviate a little bit
from the network lanes, which might be the result of the expansion
of the magnetic structures in the network, since the absorption is
mainly occurring at higher layers (upper chromosphere and TR). We
have to mention that the above result is only a statistical
correlation, since some weakly and strongly absorbed emissions are
still found at internetwork and network locations, respectively. And
one has to bear in mind that the effect of profile asymmetry can not
be excluded in the calculation of the ratio $R_{br2c}$. For
instance, this ratio will have a similar value for, e.g., a
symmetric profile with two horns of equal radiance, and an
asymmetric profile with a stronger horn on one side and a weaker
horn on the other side. But it is not straightforward to say if they
are equally reversed. Also, our method may not be appropriate for
some noisy profiles and some profiles without obvious central dips.
This effect can be neglected for \lya, but it has an impact on the
results for \lyb~ because of its weaker emission and absorption.
However, by checking individual profiles, we confirmed the
correlation shown in Fig.~\ref{fig.3}.

Our result that the Lyman line emissions are more strongly absorbed
in the internetwork as compared to the network seems to be related
with the different magnetic structures in the two regions. There is
no doubt that most of the \lya~and \lyb~emissions come from the
chromosphere and lower TR \citep{Fontenla88}. Network is the
location where magnetic flux converges, and from which magnetic
funnels (which might be related to solar wind, or just be the legs
of trans-network loops) originate \citep{Peter01}. In the network,
loops of different sizes and funnels are crowded in the chromosphere
and transition region, and the Lyman line emissions originate from
the outskirts of these magnetic structures. However, in the
internetwork region, only low-lying loops are present and most of
the Lyman line emission sources are located at a much lower height
as compared to the network. This scenario leads to an enhanced
opacity in the internetwork and thus the Lyman radiation penetrating
the upper layers will be more strongly absorbed in the internetwork
than the network.

However, we should mention that possible differences of the density
and temperature gradient in the two regions might also play a role
in the process of absorption. We should not exclude these
possibilities. To fully understand our observational result, new
detailed atmospheric models and radiative transfer calculations will
by required.

Our radiance data also allow us to test the relationship between the
central \lya~spectral radiance and the total \lya~line radiance
published by \citet{Emerich05}. The resulting scatter plots (not
shown here) suggest an almost linear relationship and a ratio of the
central spectral radiance to full line radiance of 9.43~$\pm$~0.04.
Our value is lower than the \textit{OSO}-5 result
\citep{Vidal-Madjar75}, and in closer agreement with the previous
irradiance results of \citet{Emerich05} and \citet{Lemaire05}.

\section{Ratio between \lya~and \lyb~intensities}

\begin{figure*}
\sidecaption
\includegraphics[width=12.5cm]{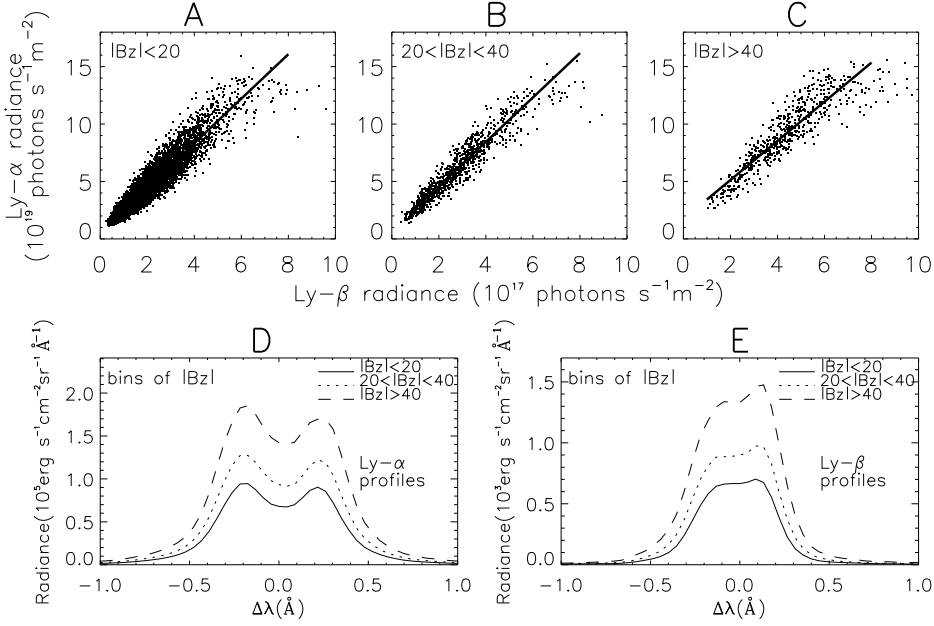}
\caption{Upper panels A to C: scatter plot of the relationship
between the \lya~and \lyb~radiances in three bins of the strength of
the vertical component of the magnetic field. The thick lines are
linear fits to the clouds of dots. Lower panels D and E: averaged
\lya~(left) and \lyb~(right) profiles in different line style
corresponding to the three bins of $\vert B_z \vert$.} \label{fig.4}
\end{figure*}

\begin {table*}[]
\caption[]{The observed ratios between \lya~and \lyb~photon
radiances.} \label{table1} \centering
\begin {tabular}{cccc}
\hline\hline
reference & target  & ratio & instrument \\
\hline
\citet{Vernazza78} & quiet Sun  & 102 & ATM UV Spectrometer\\
\citet{Vernazza78} & coronal hole  & 108 & ATM UV Spectrometer\\
\citet{Vernazza78} & active region  & 42 & ATM UV Spectrometer\\
\citet{Lemaire78} & full Sun  & 79 & LPSP/OSO 8\\
\citet{Vial82} & prominence   & 77 & LPSP/OSO 8\\
\citet{Lemaire05} & full Sun  & 155 & SUMER/SOHO\\
\citet{Vial07} & prominence   & 114-217 & SUMER/SOHO\\
\citet{Curdt08} & quiet Sun & 188 & SUMER/SOHO\\
Current paper & quiet Sun  & 212-230 & SUMER/SOHO\\
\hline
\end {tabular}
\end {table*}

One advantage of this observation over the last one is that we have
simultaneous high-rate MDI magnetograms. Thus we can sort all the
profiles according to the absolute values of the vertical magnetic
field strength, $\vert B_z \vert$, in order to find whether there is
a relationship between the magnetic field and the profile asymmetry.
We defined two thresholds, 20~Gauss and 40~Gauss, and averaged the
profiles of \lya~and \lyb~in three bins. The results are presented
in Fig.~\ref{fig.4}. A scatter plot of the relationship between the
\lya~and \lyb~radiances in each bin is also presented in that
figure.

In our previous study \citep{Curdt08}, we reported that the average
radiances of the two lines yield a \lya/\lyb~photon ratio of
$\approx188$. This result was surprising, since it significantly
deviates from values reported by \citet{Vial82} and
\citet{Vernazza78}. Here, again we applied a linear fit to a scatter
plot, and found values very similar to those obtained in the
sit-and-stare study. For three different magnetic field bins (with
increasing magnetic field strength) we found photon ratios of
$193.6\pm0.8$, $193.1\pm2.0$, and $170.3\pm3.2$, respectively.
However, from Fig.~\ref{fig.4} we can infer that some dots with high
radiances significantly deviate from the fitting line, which
indicates that the ratio should be smaller in very bright-emission
regions. We calculated the median value of the ratio of \lya/\lyb,
and found that the ratio is 229.575, 222.805 and 212.737 in the
three bins. It seems that there is a weak declining trend of the
line ratio with increasing magnetic field strength. However, the
deviation in regions with very bright emission could also be due to
unidentified saturation effect in the detection system. Thus the
insignificant difference among the three values seems to suggest
that the ratio of \lya~and \lyb~ intensities is independent of
magnetic field strength.

Recently, and also based on a SUMER observation, \citet{Vial07}
obtained varying \lya/\lyb~radiance (in the energy unit) ratios of
96, 183, and 181 in different parts of a prominence. They were
explained as relating to the fine structures in prominences. In view
of the high quality of our new data set, we are now confident that
these high values of the photon ratio are reliable within an
estimated 10\% uncertainty. Table~\ref{table1} lists the observed
ratios between \lya~and \lyb~photon radiances. Some previous results
using energy units have been converted into photon units, by
multiplying a factor of 1216/1025.

\section{Asymmetry of the profiles}

Fig.~\ref{fig.4}D\&E reveal clearly that the profile asymmetry has a
dependence on the magnetic field strength, and the asymmetries of
the average \lya~and \lyb~profiles are opposite. \citet{Curdt08}
found that the blue-horn asymmetry of \lya~and red-horn asymmetry of
\lyb~ tend to be more prominent at locations where significant
downflows are present in the transition region. Our result is
consistent with the Dopplershift-to-asymmetry relationship in
\citet{Curdt08}, since larger red shift is usually found in the
magnetic network. Thus we have confirmed that the averaged profile
is stronger in the blue horn for \lya, and stronger in the red horn
for \lyb.

In our previous paper \citep{Curdt08}, we found that there is a
relationship between the profile asymmetry and the Doppler shift in
the TR. We also found that the averaged profiles of \lya~and \lyb~
have opposite asymmetries. However, in the observation on July 2 of
2008, SUMER was operated in sit and stare mode so that only spectra
emitted from a narrow slice of the Sun were recorded. In the current
observation, we scanned a region which included several network
cells. We calculated the Doppler shifts of Si\,{\sc{iii}} and
O\,{\sc{vi}} by applying a single Gaussian fitting to the
corresponding line profiles. Using a similar analysis method, we
confirmed that the Dopplershift-to-asymmetry relationship also
exists in this new data set. Since the results are basically the
same, there is no need to present them here.

We have to mention that there is no strict (pixel-to-pixel)
correlation between the blue-horn asymmetry in \lya~and red-horn
asymmetry in \lyb. At some locations the \lya~and \lyb~profiles have
the same sign of asymmetry, but may have opposite signs of asymmetry
at other locations. This finding seems to indicate that different
processes may be responsible in both cases and is consistent with
the prediction of a recent prominence model \citep{Gunar08}, in
which the authors included both effects of Doppler shift and opacity
to calculate synthetic line profiles. We also find that most \lya~
profiles are stronger in the blue horn, while most \lyb~profiles are
stronger in the red horn.

\begin{figure*}
\sidecaption
\includegraphics[width=12.5cm]{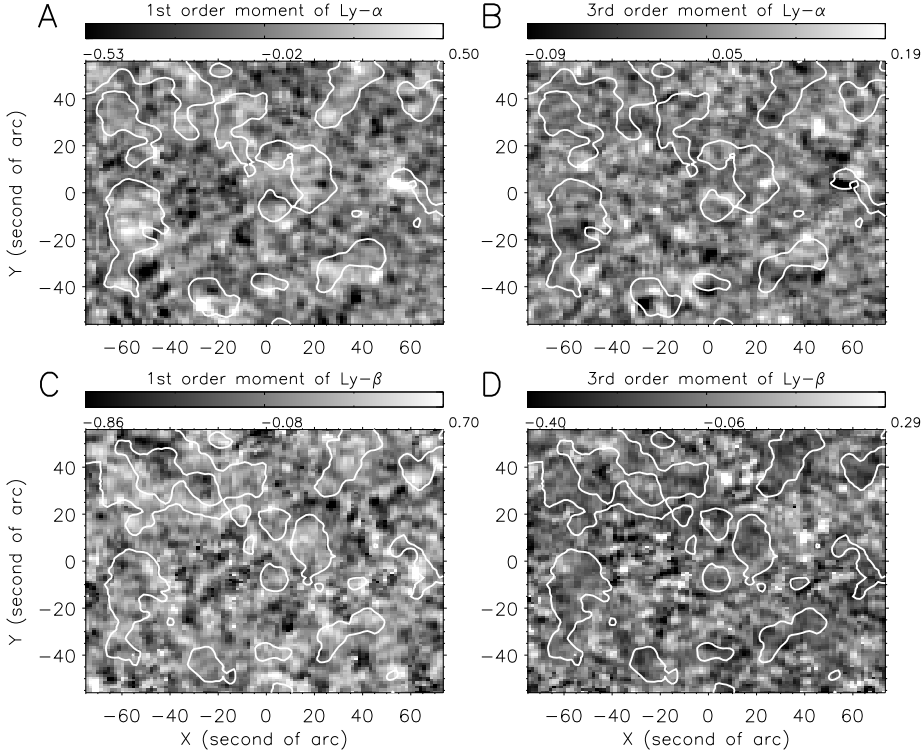}
\caption{Maps of the 1$^{st}$ (A, C) and 3$^{rd}$ (B, D) order
moments, as described in the text, of the \lya~ and \lyb~line
profiles, respectively. The white contours (enclosing the top 25\%
values) represent the brightest parts on the corresponding radiance
maps A and B of Fig.~\ref{fig.1}.} \label{fig.5}
\end{figure*}

\begin{figure*}
\sidecaption
\includegraphics[width=12.5cm]{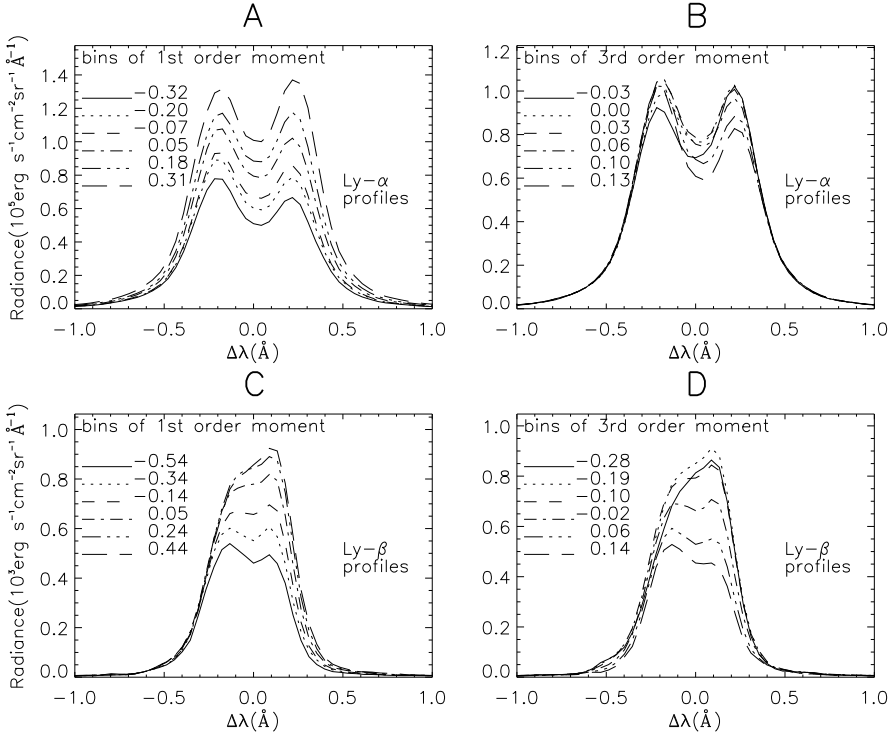}
\caption{Various averaged \lya~(upper panels A and B) and
\lyb~(lower panels C and D) profiles in bins of each moment. The
level of each bin is also shown in the plots, and the corresponding
profiles are represented by different line styles.} \label{fig.6}
\end{figure*}

To study the asymmetry of the Lyman line profiles quantitatively, we
calculated the 1$^{st}$ and 3$^{rd}$ order moments of the profiles.

\subsection{Moments of the line profiles}

The calculation of the moments of a Lyman line profile allows one to
characterize it quantitatively and to determine numbers for its
relative shift, asymmetry, peakness, and so on. Thus, one can obtain
valuable information on the state and dynamics of the solar
atmosphere. Following the definition of \citet{Gustafsson06}, we
computed the following brightness-weighted statistical moments:

\begin{equation}
\emph{mean wavelength:
\Large{$\;\lambda_{c}=\frac{\sum~(I_{i}-B_{i})\lambda_{i}}{\sum~(I_{i}-B_{i})}$}}
\label{equation1},
\end{equation}
\begin{equation}
\emph{standard deviation:
\Large{$\;\sigma=\sqrt{\frac{\sum~(I_{i}-B_{i})(\lambda_{i}-\lambda_{c})^{2}}{\sum~(I_{i}-B_{i})}}$}}
\label{equation2},
\end{equation}
\begin{equation}
\emph{skewness:
\Large{$\;\frac{\sum~(I_{i}-B_{i})(\lambda_{i}-\lambda_{c})^{3}}{\sigma^{3}\sum~(I_{i}-B_{i})}$}}
\label{equation3},
\end{equation}

where $I_{i}$, $B_{i}$, and $\lambda_{i}$ represent, respectively,
the spectral radiance, background, and wavelength (here in the unit
of a spectral pixel $i$, which is about 42~m{\AA}). For the sake of
simplicity, we assumed a constant continuum background, with the
level being given by the lowest spectral radiance of the profile.
After calculating the moments for each profile, we constructed a map
of each moment of \lya~and \lyb. Here we only present the maps of
the 1$^{st}$ and 3$^{rd}$ order moments of the two Lyman lines in
Fig.~\ref{fig.5}. The white contours (enclosing the top 25\% values)
on the maps represent the brightest parts on the corresponding
radiance map of Fig.~\ref{fig.1}, which clearly outline the network
pattern.

The 1$^{st}$-order moment represents the mean spectral position of
the profile. Since we are mainly interested in the relative
position, we subtracted the average (over the entire studied region)
mean position from the mean position of each profile. The resulting
value will be referred to as the 1$^{st}$-order moment in our later
discussion. It corresponds to the Doppler shift in the case of a
Gaussian-shaped profile. However, in the case of \lya~and \lyb, not
only the Doppler shift of the line, but also the asymmetry of the
line profile can contribute to the 1$^{st}$-order moment.

The 2$^{nd}$-order moment is the standard deviation of the profile.
In the case of \lya~and \lyb, it includes information on both the
line width and the self-reversal depth. However, a narrow profile
with a deep self-reversal and a broad profile with a shallow
self-reversal may have a comparable value of the 2$^{nd}$-order
moment. Since we wanted to avoid putting these two different kinds
of profiles into the same category, we did not analyse this moment
in our study.

The 3$^{rd}$-order moment, the skewness, is a measure of the
asymmetry of the line profiles. This parameter seems to be a good
candidate to study the asymmetry of \lya~and \lyb~profiles. However,
both \lya~and \lyb~are blended with a He\,{\sc{ii}} line in the blue
wing. This blend will definitely influence the calculation of the
3$^{rd}$-order moments. The problem for \lyb~is even worse, since
the line is much narrower and weaker than \lya~so that the
He\,{\sc{ii}} blend contributes more to the skewness of \lyb. And
even more, there is another O\,{\sc{i}} blend in the center part of
the \lyb~profile. The mixture effect of the blends can be seen in
the values calculated from Eq~(\ref{equation3}). Therefore, these
values were adjusted by adding 0.05 for \lya~and subtracting 0.24
for \lyb, to make those profiles with equal-height horns correspond
to a zero value of skewness.

We have sorted the profiles by each moment and defined six bins
which were equally spaced in the moment. The average profiles in
each bin are shown in Fig.~\ref{fig.6}. Again, only the results of
the 1$^{st}$ and 3$^{rd}$ order moments, which we are mainly
interested in, are presented. The averaged profiles exhibit obvious
signatures of asymmetry, which will be further investigated in the
next two sections.

\subsection{Discussion}

It is not easy to find a parameter which can purely reflect the
asymmetry of the Lyman line profiles. However, from Fig.~\ref{fig.6}
the 1$^{st}$- and 3$^{rd}$-order moments seem to reveal some
information, although not purely, on the line asymmetry. As
mentioned previously, both the shift of the entire profile and the
asymmetry of the profile can contribute to the 1$^{st}$-order moment
of \lya~and \lyb. A red-shifted profile with a stronger red wing
will definitely lead to a larger value of the 1$^{st}$-order moment.
In contrast, a blue-shifted profile with a stronger blue wing should
correspond to a smaller value of the 1$^{st}$-order moment. From
panels A and C of Fig.~\ref{fig.5} we find that in statistical
sense, both \lya~and \lyb~have a larger value of their
1$^{st}$-order moment in the network than in the internetwork. This
behaviour indicates that the average network-related profile is
either largely red-shifted or strongly enhanced in the red wing. As
mentioned previously, the 3$^{rd}$-order moment, i.e., the skewness
of the \lya~and \lyb~ profile, is influenced by the blends. However,
by inspection of Fig.~\ref{fig.6} we find that the asymmetries of
the Lyman line profiles still can somehow be reflected in the values
of the 3$^{rd}$-order moment. A negative/positive value of this
parameter corresponds to a profile with an enhanced red/blue horn.

\begin{figure*}
\resizebox{\hsize}{!}{\includegraphics{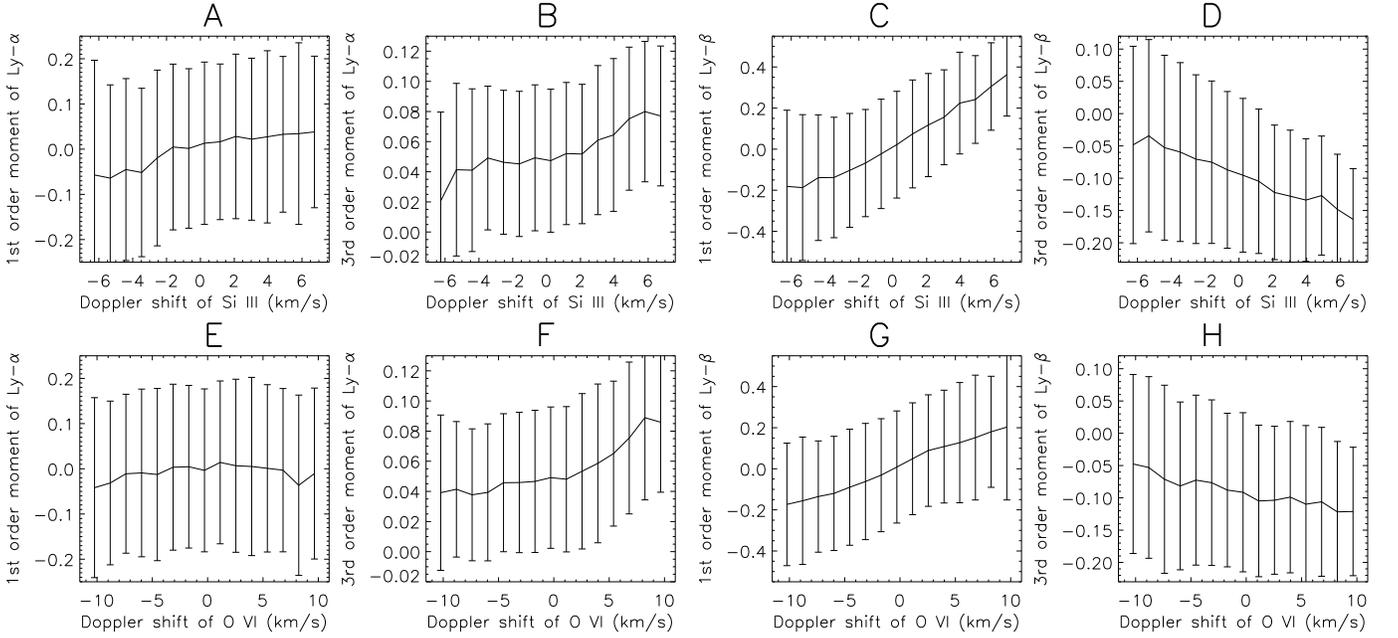}}
\caption{Dependence of the 1$^{st}$ and 3$^{rd}$-order moments on
the Doppler shifts of the Si\,{\sc{iii}} line (panels A to D) and
O\,{\sc{vi}} line (panels E to H). A positive and negative value of
the Doppler shift indicates redshift (downflow) and blueshift
(upflow), respectively. The error bar indicates the standard
deviation of the corresponding moment in each bin.} \label{fig.7}
\end{figure*}

Since the asymmetries of the Lyman line profiles are caused by a
combined effect of Doppler shift and opacity \citep{Gunar08}, we may
find that there is a relationship between the 1$^{st}$ and 3$^{rd}$
order moments and the Doppler shift. Here we can only calculate the
Doppler shifts of two transition region lines, Si\,{\sc{iii}} and
O\,{\sc{vi}}. In Fig.~\ref{fig.7}, we present the dependence of the
1$^{st}$ and 3$^{rd}$ order moments of \lya~and \lyb~on the Doppler
shifts of Si\,{\sc{iii}} and O\,{\sc{vi}}.

\begin{figure*}
\sidecaption
\includegraphics[width=12.5cm]{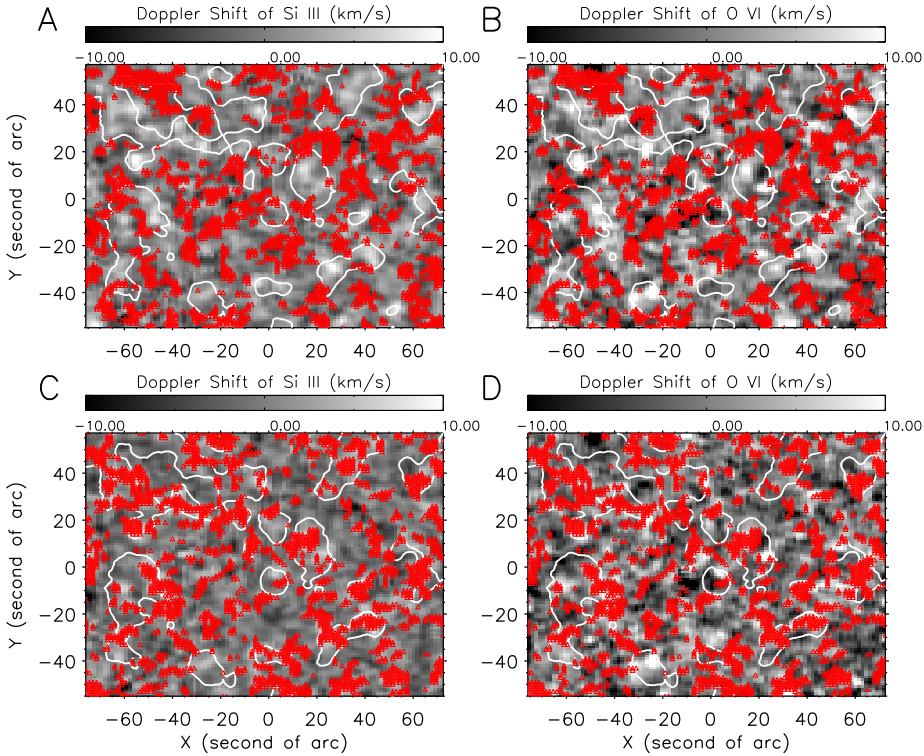}
\caption{Locations of the \lyb~profiles corresponding to the 25\%
smallest values (upper panels, A and B) and the 25\% largest values
(lower panels, C and D) of the 1$^{st}$-order moment, which are
marked by small red triangles on the grey-shaded Dopplergrams of the
Si\,{\sc{iii}} and O\,{\sc{vi}} lines. The white contours (enclosing
the top 25\% of the radiance values) represent the brightest parts
on the \lyb~radiance map.} \label{fig.8}
\end{figure*}

It is clear that the 1$^{st}$-order moment of \lyb~is positively
correlated with the Doppler shift in the TR. This relationship is
corroborated by Fig.~\ref{fig.8}, in which locations of \lyb~
profiles corresponding to the 25\% smallest values (upper panels)
and 25\% largest values (lower panels) of the 1$^{st}$-order moment
are marked by red triangles on the Dopplergrams of Si\,{\sc{iii}}/
O\,{\sc{vi}}. The white contours outline the network pattern as seen
on the \lyb~radiance map. This strong correlation can be understood
if we assume that the \lyb~profile is more redshifted and more
enhanced in the red horn in regions where larger TR downflows are
present. Panels D and H of Fig.~\ref{fig.7} reveal a clear
decreasing trend of the skewness with increasing TR Doppler shift,
which means that the red horn of \lyb~tends to be more enhanced
where the Doppler shift is larger.

To further substantiate this result, we extended the analysis on the
data set used for the spectral atlas of \citet{Curdt01} and
completed a detailed wavelength calibration for \lyb, based on
nearby fluorecence lines of neutral oxygen as wavelength standards.
These lines are radiatively excited, probably by Lyman continuum
photons via the 3s, 3d, and 4s levels of the O\,{\sc{i}} triplet
system. We assume that these optically thin O\,{\sc{i}} lines emerge
from the lower chromosphere \citep{Avrett08} and are at rest, at
least within the required uncertainties. We could reproduce the
average line profile with a multigauss fit of a broad Gaussian line
($\approx$0.3 A FWHM) with a redshift of 5~km/s, the He\,{\sc{ii}}
Balmer line in the blue wing, a much narrower Gaussian absorption
profile ($\approx$0.15 A FWHM), and the O\,{\sc{i}} lines as
wavelength standards. Because of the lack of wavelength standards
and the much stronger absorption, this method cannot be used for
\lya.

This redshift of 5~km/s is comparable to the average redshift of a
typical TR line \citep{Brekke97,Warren97,Chae98,Xia03}. In addition,
following the method described in \citet{Curdt08a}, we divided the
average network profile by the average internetwork profile and
obtained the network contrast profile for \lyb~(not shown here). We
found that there is a clear peak of this network contrast profile on
the red side of the line profile, which indicates that the \lyb~line
is redshifted. Such a behaviour is similar to typical TR lines and a
natural result of the well-known brightness-Doppler-shift
relationship. Although the hydrogen Lyman lines are mainly formed in
the chromosphere and lower TR, since hydrogen is the most abundant
element in the solar atmosphere, there should still be some emission
sources of its resonance lines in the middle and upper TR. \lyb~is
optically thinner than \lya, and we can see substantial \lyb~
emission from the TR. That is why the behaviour of \lyb~is similar
to typical TR lines.

Panels B and F of Fig.~\ref{fig.7} reveal a clear increasing trend
of the skewness with increasing TR Doppler shift, which means that
the blue horn of \lya~tends to be more enhanced if the TR redshift
is larger. This is consistent with the result of \citet{Curdt08}.
From panels A and E of Fig.~\ref{fig.7} we find that the
1$^{st}$-order moment of \lya~tends to slightly increase or stay the
same with increasing TR Doppler shift. As we mentioned previously,
both the Doppler shift of the whole profile and the profile
asymmetry can contribute to the 1$^{st}$-order moment. Since the
enhanced blue horn will reduce the moment, the only possibility is
that \lya~is more redshifted in regions where larger downflows are
present.

The large scatters in Fig.~\ref{fig.7} suggest that besides the TR
flows, there should be other effects which can influence the
asymmetry of the Lyman line profiles. We have to mention that the
explanation of the asymmetries of the \lya~and \lyb~profiles will
require intricate modelling rather than simple imagination. Their
different asymmetries arise from the motions of the solar atmosphere
in its different layers as well as from the different line
opacities. Further theoretical work, especially modeling, is needed
to improve our understanding of the formation of the Lyman line
profiles.

As mentioned above, energy losses through the \lya~ emission are the
most important radiative losses in the lower TR \citep{Fontenla88}.
Thus the \lya~line should be important for the diagnostic of the
magnetic and plasma properties in the lower TR. It is still under
debate if the TR emission originates from a thin thermal interface
connecting the chromosphere and corona
\citep{Gabriel1976,Dowdy1986,Peter01}, or from isolated unresolved
tiny loops \citep{Feldman1983,Feldman1987,Feldman1994}. Recently,
\cite{Patsourakos2007} reported the first spatially resolved
observations of subarcsecond-scale (0.3$^{\prime\prime}$) loop-like
structures seen in the \lya~line, as observed by the Very High
Angular Resolution Ultraviolet Telescope (VAULT). As mentioned in
that paper, flows can have distinct spectroscopic signatures, which
can help to distinguish between the two possibilities of the TR
emission origin. However, the signature of flows in the corona is
missing in the SUMER observation. Moreover, the spatial resolution
of our SUMER observation is still not high enough. Thus we can still
not make a final conclusion on the origin of the TR emission. We
think that only through a combined analysis of high-resolution
(subarcsecond) magnetic field, EUV imaging, and spectroscopic
observations can further substantial progress be expected.

\section{Summary and conclusion}

We have presented new results from a quasi-simultaneous observation
of the profiles of \lya~and \lyb. We find that the self-absorption
in or around network is reduced, and in the internetwork region the
emissions tend to be more strongly absorbed in the central part of
the profiles. We suggest that the different magnetic structures in
the two regions might be responsible for this result. The outskirts
of the chromospheric and transition-region magnetic structures,
where the Lyman line emissions originate, are higher in the network
as compared to the internetwork.

Using the simultaneously measured photospheric magnetic field, we
could determine the ratio of \lya~and \lyb~ intensities for
different bins of magnetic field strength. Our result seems to
suggest that the ratio is independent of the magnetic field
strength.

A rough inspection of the line profiles suggests that most
\lya~profiles are stronger in their blue horn, while most \lyb~
profiles are stronger in their red horn. But the opposite
asymmetries of \lya~and \lyb~are not strictly (on a pixel-to-pixel
level) correlated.

The skewness of the profiles reveals clearly that the \lya~and
\lyb~profiles are more enhanced in the blue and red horns,
respectively, if larger TR downflows are present. The first-order
moment of \lyb~is strongly correlated with the Doppler shift of
Si\,{\sc{iii}} and O\,{\sc{vi}}. This result can be understood if
the profiles are more redshifted in regions where larger TR
downflows are present. The correlation is much weaker for \lya,
which indicates that the \lya~profiles should also be more
redshifted if larger TR downflows are present. Through a multi-Gauss
decomposition of \lyb, we found that \lyb~is redshifted on average.

The opposite asymmetries of the average profiles of \lya~and \lyb~
are presumably caused by the combined effects of flows in various
layers of the solar atmosphere and opacity difference of the two
lines. A mechanism for line formation can not be simply imagined but
must be thoroughly devised and further investigated by help of
models.

\begin{acknowledgements}
The SUMER project is financially supported by DLR, CNES, NASA, and
the ESA PRODEX Programme (Swiss contribution). SUMER and MDI are
instruments onboard {\it SOHO}, a mission operated by ESA and NASA.
We thank the teams of SUMER and MDI for the spectroscopic and
magnetic field data. We thank Dr. L.-D. Xia for the helpful
discussion. We also thank the referee for his/her careful reading of
the paper and for the comments and suggestions.

Hui Tian is supported by the IMPRS graduate school run jointly by
the Max Planck Society and the Universities of G\"ottingen and
Braunschweig. The work of Hui Tian's group at PKU is supported by
the National Natural Science Foundation of China (NSFC) under
contract 40874090.
\end{acknowledgements}

\end{document}